\documentclass{article}
\usepackage{graphicx}
\usepackage{amsmath}
\usepackage{amsfonts}
\usepackage{amssymb}

\begin{document}

\title{A feasible quantum communication complexity protocol}
\author{Ernesto F. Galv\~{a}o\\Centre for Quantum Computation, Clarendon Laboratory,\\Univ. of Oxford, Oxford OX1 3PU U.K.}
\maketitle
\begin{abstract}
I show that a simple multi-party communication task can be performed more
efficiently with quantum communication than with classical communication, even
with low detection efficiency $\eta$. The task is a communication complexity
problem in which distant parties need to compute a function of the distributed
inputs, while minimizing the amount of communication between them. A realistic
quantum optical setup is suggested that can demonstrate a five-party quantum
protocol with higher-than-classical performance, provided $\eta>0.33$.
\end{abstract}

In theory, quantum communication is better than classical communication.
Experimentalists, on the other hand, know that even the simplest quantum
communication protocols involve inefficiencies in state preparation,
manipulation and measurement. It is, therefore, important to study sufficient
experimental conditions for unambiguous demonstration of the advantages of
quantum communication. Some tasks are only possible with quantum
communication, such as unconditionally secure cryptographic key distribution
\cite{Bennett B 84, Ekert 91, Bennett 92}. Many authors have analyzed the
experimental requirements for the security of these protocols \cite{Lo C 99,
Shor P 00, Aschauer B 00}. For other tasks quantum communication offers an
improvement of efficiency, and such is the case of communication complexity
problems \cite{Yao 79, Kushilevitz N 97}, one of which will be analyzed in
this paper. In these problems many distant parties need to compute a function
of the distributed inputs, while trying to minimize the amount of
communication between them. This abstract problem has numerous practical
applications, for example in computer networks, VLSI circuits and data
structures (see \cite{Kushilevitz N 97} for a survey of the field).

Quantum mechanics can enhance the performance of communication complexity
protocols in two different ways \cite{Brassard 01}. The first approach is the
\textit{entanglement-based} model of communication complexity \cite{Cleve B
97, Grover 97, Buhrman CvD 97, Buhrman vDHT 99}, where in addition to the
classical communication we allow the parties to do measurements on previously
shared multi-party entangled states. Experimental requirements for some
protocols of this kind have been studied in \cite{vDam 99, Galvao 00}, and it
turns out that the high detection efficiency needed could be achieved in ion
trap experiments \cite{Rowe KMSIMW 01}. The second way to obtain a genuine
quantum advantage is to allow the parties to exchange qubits instead of
classical bits \cite{Yao 93, Buhrman CW 98, Cleve vDNT 99, Raz 99}. That such
a \textit{quantum communication} model may be superior to the classical case
is surprising, given the results of Holevo \cite{Holevo 73} and Nielsen
\cite{Cleve vDNT 99, Nielsen 98} that state that no more than $n$ bits of
expected information can be transmitted by $n$ qubits, if the parties start
off unentangled. Despite the many theoretical results obtained by different
authors \cite{Brassard 01}, to date no experiment has been performed to
demonstrate the superiority of quantum communication for this kind of
distributed computation task. In this paper I propose a feasible quantum
optical experiment which implements a quantum protocol with
higher-than-classical performance for a specific communication complexity
task. The quantum advantage is shown to arise from the use of a quantum phase
to encode information. A realistic estimate of all experimental limitations
shows that it is sufficient to have a single-photon detection efficiency
$\eta\gtrsim0.33$ for the quantum protocol to outperform any classical
protocol for the same problem.

The communication complexity problem we will tackle is the \textit{Modulo-4
Sum }problem defined for three parties by Buhrman, Cleve and van Dam
\cite{Buhrman CvD 97}, and later generalized to $N$ parties ($N\geq3$) in
\cite{Buhrman vDHT 99}. The problem can be stated as follows. Each party
$P_{i}$ receives a two-bit string input $x_{i}$, subject to the constraint:%

\begin{equation}
\left(  \sum_{i=1}^{N}x_{i}\right)  \operatorname{mod}2=0. \label{constraint}%
\end{equation}
The strings are chosen randomly with an uniform probability distribution among
those combinations that satisfy eq. \ref{constraint} above. After some
communication between the parties, one of them (say the last one $P_{N}$) must
compute the value of the Boolean function%

\begin{equation}
F(\overrightarrow{x})=\frac{1}{2}\left[  \left(  \sum_{i=1}^{N}x_{i}\right)
\operatorname{mod}4\right]  .
\end{equation}
In other words, each party is given a number $x_{i}\in\{0,1,2,3\}$, subject to
the constraint that the sum of all $x_{i}$ is even. After some communication
the last party must decide whether the sum modulo-$4$ is equal to $0$ or $2$.

References \cite{Buhrman CvD 97, Buhrman vDHT 99} dealt with this problem in
the entanglement-based model of communication complexity, showing that the
amount of classical communication necessary to compute $F$ (on inputs
constrained by eq. \ref{constraint}) can be decreased if the parties are
allowed to do local measurements on $N$-party Greenberger-Horne-Zeilinger
(GHZ) states%
\begin{equation}
\left|  GHZ\right\rangle =\frac{1}{\sqrt{2}}\left(  \left|  0_{1}0_{2}%
\cdots0_{N}\right\rangle +e^{i\phi}\left|  1_{1}1_{2}\cdots1_{N}\right\rangle
\right)  . \label{GHZ state}%
\end{equation}

When considering the quantum communication model, we must limit the amount of
bits (qubits) to be exchanged between the parties and compare the success
rates obtained by the optimal classical and the quantum protocols. The
criterion for a successful demonstration of better-than-classical
communication is simple: we just need to obtain an experimental quantum
success rate which is better than that of the optimal classical protocol.

Let us limit the amount of communication to $(N-1)$ bits (or qubits). Another
constraint we impose is that the communication must be \textit{sequential}, in
which party $P_{1}$ can only send information to party $P_{2}$, who in turn
can only send a message to party $P_{3}$ and so on until party $P_{N}$, who
then computes $F$. The decision to demand sequential communication is related
to the fact that the sequential quantum communication necessary to solve this
problem can be conveniently realized by sending a single photon through a
series of optical elements representing the parties.

First, let us obtain the optimal classical success rate for the Modulo-4 Sum
problem, with only $(N-1)$ bits of sequential classical communication. We
start by noting that if one of the parties (say party $P_{j}$) sends no
information to party $P_{j+1}$, then party $P_{N}$ cannot compute $F$
correctly with probability $p_{c}>1/2$. This is so because such a break in the
communication flow would leave party $P_{N}$ with no information about the
numbers $x_{1,}x_{2},...,x_{j}$, and there are as many allowed $j$-tuples
$(x_{1},x_{2,}...x_{j})$ resulting in $F(\overrightarrow{x})=1$ as in
$F(\overrightarrow{x})=0$. Therefore, in order to obtain a performance which
is better than a random guess, each party $P_{j}$ must send exactly one bit to
the next party $P_{j+1}$ .

For the moment let us consider only deterministic protocols. The first party
$P_{1}$ has access only to her two-bit string $x_{1}$, and so can choose
between $2^{4}$ protocols. These can be represented by the four-bit string
$prot_{1}$, whose $n^{th}$ ($n=0,1,2,3$) bit encodes the message $m_{1}$ to be
sent to $P_{2}$ if $x_{1}=n$. The other parties $P_{j}$ ($j=2,...,N-1$) can
choose among $2^{8}$ protocols that take into consideration both $x_{j}$ and
the message $m_{j-1}$ received from the previous party. Each of these
protocols can be represented by an 8-bit string $prot_{j}$, whose $n^{th}$
($n=0,1,...,7$) bit encodes the message to be sent when $x_{j}+2m_{j-1}=n$.

Each possible deterministic protocol can then be represented by the
$(N-1)$-tuple $\overrightarrow{prot}=(prot_{1},prot_{2},...,prot_{N-1})$.
Finding the probability of success of a given protocol $\overrightarrow{prot}$
is a straightforward computation. We start by producing a list of all possible
input data $\{x_{1},x_{2},...,x_{N-1}\}$ compatible with $x_{N}=0$, computing
the messages $m_{N-1}$ corresponding to each, and finding the fraction of
cases in which $P_{N}$'s most likely guess about $F$ would in fact be correct.
This is repeated for $x_{N}=1,2$ and $3$, and the results averaged to obtain
the overall probability of success $p_{c}$. The optimal deterministic protocol
can then be found by a computer search over all $2^{4}(2^{8})^{N-2}%
=2^{(8N-12)}$ protocols.

For number of parties $N=3,4$ and $5$ I obtained the optimal classical
probability of success%

\begin{align}
p_{c}^{N=3}  &  =3/4,\\
p_{c}^{N=4}  &  =3/4,\nonumber\\
p_{c}^{N=5}  &  =5/8.\nonumber
\end{align}
A limited search over protocols for larger number of parties yields some lower
bounds for $p_{c}$:%

\begin{align*}
p_{c}^{N=6}  &  \geq5/8,\\
p_{c}^{N=7}  &  \geq9/16,\\
p_{c}^{N=8}  &  \geq9/16.
\end{align*}
Since $p_{c}$ is a non-increasing function of $N$, the result for $N=6$ is
actually an equality: $p_{c}^{N=6}=5/8$. The optimal $p_{c}$ for $N=3,4,5$ and
$6$ is attained by many protocols, for example the one consisting of
$prot_{0}=0011$ and all the other $prot_{j}=01011010$. The same protocol
yields the lower bounds for the optimal probabilities of success presented
above for $N=7$ and $8$. Checking that these lower bounds are tight would
involve a very long exaustive search over all protocols. For the purpose of
comparison with the quantum protocol given below, it would be desirable to
obtain at least an analytical upper bound for $p_{c}^{N}$ that decreases with
$N$. Unfortunately I could not prove such a general result, despite the
symmetries of the problem.

Up to now we have been computing the probability of success for deterministic
protocols. In a probabilistic protocol, each party $P_{j}$ implements her own
protocol by probabilistically picking a deterministic protocol $prot_{j}$ from
some set of protocols, according to probabilities obtained from a list of
random numbers. Since this list of numbers could have been shared beforehand
between the parties, the last party $P_{N}$ can know exactly which protocols
were chosen by each of the other parties for each run of the probabilistic
protocol. This means that each run of the probabilistic protocol is
effectively a deterministic one, with a probability of success bounded by the
optimal deterministic $p_{c}$ derived above. The relation between
deterministic and probabilistic protocols for communication complexity tasks
is further discussed in chapter 3 of the book by Kushilevitz and Nisan
\cite{Kushilevitz N 97}.

We have seen that the Modulo-4 Sum problem gets harder and harder to solve
classically, as the number of parties increases. There is, however, a simple
quantum protocol with sequential qubit communication that has a probability of
success $p_{q}=1$ \textit{independently} of the number of parties involved.
The idea is to start with the qubit in state%

\[
\left|  \psi_{i}\right\rangle =\frac{1}{\sqrt{2}}\left(  \left|
0\right\rangle +\left|  1\right\rangle \right)
\]
and send it flying by all the parties, from first to last. Each party needs
only act upon the qubit with a phase operator $\phi(x_{j})$, defined as%

\begin{equation}
\phi(x_{j})=\left\{
\begin{array}
[c]{c}%
\left|  0\right\rangle \rightarrow\left|  0\right\rangle \\
\left|  1\right\rangle \rightarrow e^{i\frac{\pi}{2}x_{j}}\left|
1\right\rangle
\end{array}
\right.  ,x_{j}=\{0,1,2,3\}. \label{phase op}%
\end{equation}
After going through the $N$ phase operations the qubit state will be%

\[
\left|  \psi_{f}\right\rangle =\frac{1}{\sqrt{2}}\left(  \left|
0\right\rangle +(-1)^{F(\overrightarrow{x})}\left|  1\right\rangle \right)  ,
\]
due to the constraint \ref{constraint} on the possible inputs $x_{j}$. The
last party can then measure $\left|  \psi_{f}\right\rangle $ in the
$\{\frac{1}{\sqrt{2}}(\left|  0\right\rangle +\left|  1\right\rangle
),\frac{1}{\sqrt{2}}(\left|  0\right\rangle -\left|  1\right\rangle \}$ basis,
obtaining $F$ with probability $p_{q}=1$.

The protocol above is an adaptation of the entanglement-based protocol
presented in \cite{Buhrman vDHT 99} to the qubit-communication setting. In the
entanglement-based protocol each party performs a phase operation and
measurement on her qubit of the $N$-party GHZ state they share. The value of
the function $F$ is encoded in the quantum phase $\phi$ (see eq. \ref{GHZ
state}), by individual phase shifts applied by each party on her particle.The
last party $P_{N}$ obtains the value of $F$ from the results of the $N$
measurements (hers plus the $N-1$ broadcast to her by the other parties). The
probability of success is $p_{q}=1$ only when all the $N$ detections are
successful, hence the high detection efficiencies required for a
higher-than-classical performance \cite{Galvao 00}. Here we obtain the same
performance by using the phase of a \textit{single} qubit to acquire
information on $F$ as it flies by the parties towards the last party $P_{N}$,
where a single detection reveals the result.

The detection efficiency $\eta$ must still be taken into account, as it lowers
the probability of success of the quantum protocol. For the moment, let us
assume that the only limitation in implementing the protocol is $\eta<1$ (we
will deal with the more realistic case below). In case of a successful
detection (which occurs with probability $\eta$) the probability of success is
equal to one. In case the detection fails (probability $1-\eta$), the last
party $P_{N}$ has to make a random guess about the value of $F$, succeeding
only with probability $1/2$. Thus for a higher-than-classical performance we
need to implement the quantum protocol with a detection efficiency $\eta$ such that%

\begin{equation}
\eta+(1-\eta)\frac{1}{2}>p_{c}.
\end{equation}
Thus, it is sufficient to have $\eta>2p_{c}-1$. We have seen that the optimal
classical protocol for $N=5$ parties has a success rate $p_{c}^{N=5}=5/8$, and
therefore can be beaten by the quantum protocol if the detection efficiency
$\eta>0.25$, in the absence of other experimental losses.

For a more realistic grasp of the experimental difficulties, let us examine a
simple quantum optical setup that implements the quantum protocol for this
problem. The flying qubit is encoded in the polarization state of a single
photon. For a fair comparison with the classical protocol, it is important to
allow only a single photon per run to pass by the parties and arrive at
$P_{N}$. One way to achieve this is to use a parametric down conversion
crystal pumped by a laser. Detection of one of the twin photons generated can
then be used as a trigger to let the second photon go towards the parties. For
the triggering mechanism to work we need to introduce a delay for the second
photon, which can be easily achieved by coupling it to a few meters of optical
fiber. Upon detection of the first photon, the second photon is allowed to
come through the $N$ parties. Each party consists of an optical element using
birefringent materials to perform the phase shift given by eq. \ref{phase op}.
In the end, the last party $P_{N}$ must also detect the photon in the proper basis.

Such a setup has other imperfections that must be considered, besides the
limited detection efficiency $\eta$. The first is the finite transmissivity
$t$ of the combination of $N$ birefringent plates used to introduce the phase
shifts $\phi(x_{j})$. Another problem is the fraction $\mu$ of detected events
which are due to detector dark counts. Finally, even if the detected photon is
a signal photon, the success rate $s$ of the quantum protocol can be less than
perfect, because of imperfections and misalignment of the optical elements
that produce the initial state, introduce the phase shifts and measure the
final polarization. Taking all these limitations into account, for a
higher-than-classical probability of success we would need:%

\begin{equation}
p_{q}^{eff}=(1-\mu)\eta ts+[1-(1-\mu)\eta t]\frac{1}{2}>p_{c}.
\label{big ineq}%
\end{equation}
Now let us make some realistic estimates for these parameters for the protocol
with $N=5$ parties. By using quartz plates with anti-reflection coating, it is
possible to obtain transmission of a fraction $0.995$ of the incident photons
per plate, which in the case of five parties would result in $t=(0.995)^{5}%
\simeq0.975$. It is relatively straightforward to bring dark count rates below
the $1\%$ level \cite{Kurtsiefer OW 01}, so let us take $\mu=0.01$. Good
alignment of the optical elements should enable a success rate of
$s\simeq0.90$ whenever a signal photon is detected; for example, visibilities
of up to $96\%$ in simple Bell tests using entangled photons have been
reported \cite{Kurtsiefer OW 01}. Plugging these estimates for $t,\mu$ and $s$
in inequality \ref{big ineq}, we see that in order to obtain a
better-than-classical probability of success it is sufficient to have a
detection efficiency $\eta\gtrsim0.33$, which is within reach of current
technology \cite{Kurtsiefer OW 01}. It is reasonable to conjecture that the
optimal $p_{c}^{N}$ continues to decrease for $N\geq7$, in which case the
sufficient detection efficiency could be dramatically lower. In principle, one
way to calculate $p_{c}^{N}$ for $N\geq7$ is through an exhaustive search over
all deterministic protocols, as was done here for $N=3,4$ and $5$.

It is clear that essentially the same setup can be used to solve the Modulo-4
Sum problem using classical polarized light. In common with a qubit, classical
light has a continuous variable (the phase) that can be manipulated, as
opposed to classical bits that can only assume two discrete values. The
counter-intuitive quantum feature that helps in communication complexity is
the fact that even single photons still retain the continuous description of
the classical electromagnetic field. More generally, a $d$-dimensional pure
quantum state is characterized by $2(d-1)$ real parameters that can be used
for communication purposes, as opposed to the $d$ discrete states available to
a classical system of same dimensionality. Defining exactly for which
communication tasks such a different resource can be used to advantage is a
central research problem in quantum information theory.

In summary, I have shown that an experimental demonstration of a quantum
communication complexity protocol is feasible using a realistic quantum
optical setup with photon detection efficiency of at least $33\%$. By
increasing the number of parties $N$ it should be possible to reduce the
minimum detection efficiency required dramatically, provided we can compute
the corresponding optimal classical probability of success $p_{c}^{N}$. This
can in principle be achieved by the methods employed here, or possibly by
other, simpler arguments. The higher-than-classical performance of the quantum
communication protocol arises directly from the use of a quantum phase to
encode information. If implemented, this would be the first experiment to
demonstrate the superiority of quantum communication over classical
communication for distributed computation tasks.

I would like to thank Lucien Hardy for fruitful discussions and for his
support and encouragement. I also thank Antia Lamas-Linares for helpful
discussions about quantum optics experiments. I acknowledge support from the
U.K. Overseas Research Studentships scheme and from the Brazilian agency
Coordena\c{c}\~{a}o de Aperfei\c{c}oamento de Pessoal de Nivel Superior (CAPES).

\textit{Note added}: After this work was completed, it came to my attention
that the three-party qubit communication protocol has recently been discussed
in ref. \cite{Buhrman CvD 01}, which is an extended version of ref.
\cite{Buhrman CvD 97}.

\end{document}